\documentclass[twocolumn,floatfix,prb,aps,showpacs]{revtex4-2}
\usepackage{graphicx,amsmath,amssymb,color}
\usepackage{nicefrac}
\usepackage[titletoc,title]{appendix}
\usepackage[colorlinks,bookmarks=true,citecolor=blue,linkcolor=red,urlcolor=blue]{hyperref}

\begin{document}

\title{Abelian parton state for the $\nu=4/11$ fractional quantum Hall effect}
\author{Ajit C. Balram}
\affiliation{Institute of Mathematical Sciences, HBNI, CIT Campus, Chennai 600113, India}
\date{\today}

\begin{abstract} 
We consider the fractional quantum Hall effect at the filling factor $\nu=4/11$, where two independent experiments have observed a well-developed and quantized Hall plateau. We examine the Abelian state described by the ``$4\bar{2}1^{3}$" parton wave function and numerically demonstrate it to be a plausible candidate for the ground state at $\nu=4/11$. We work out the low-energy effective theory of the $4\bar{2}1^{3}$ edge and make predictions for experimentally measurable properties of the state. 
\end{abstract}

\maketitle

\section{Introduction}
The marvelous phenomena of the fractional quantum Hall effect (FQHE)~\cite{Tsui82} arise from interactions between electrons subjected to a perpendicular magnetic field and cooled to low temperatures. FQHE in the lowest Landau level (LLL) predominantly occurs at filling factors of the form $\nu=n/(2pn\pm 1)$, where $n$ and $2p$ are positive integers. These fractions can be attributed to the integer quantum Hall effect (IQHE) of composite fermions (CFs), which are bound states of an electron and even number ($2p$) of vortices~\cite{Jain89}. Almost all the FQHE phenomenology occurring in the LLL can be understood using the theory of weakly-interacting CFs~\cite{Jain07}. However, in the range of filling factors $1/3<\nu<2/5$ samples with high mobility exhibit FQHE features at fractions $\nu=4/11,~5/13,~6/17,~3/8,~3/10,~4/13$, and $5/17$~\cite{Pan03}. These states cannot be explained as IQHE states of CFs and likely arise from an FQHE of CFs themselves~\cite{Mukherjee12, Mukherjee14c, Balram16c}.  

Recently, Samkharadze \emph{et al.}~\cite{Samkharadze15b} demonstrated activated magnetotransport at $\nu=4/11$ and $5/13$, measured their gaps, and thereby fully confirmed the formation of incompressible state at these two fillings. Interestingly, they found that at the lowest temperatures, incompressibility only occurs at 4/11 and 5/13 while fractions such as 3/8 and 6/17 remain compressible: as the temperature is lowered, the minimum in the longitudinal resistance deepens at 4/11 and 5/13 while the minima at 3/8 and 6/17 go away. They did not observe any signatures of FQHE at 3/10, 4/13, and 5/17. In an independent experiment, Pan \emph{et al. }~\cite{Pan15} demonstrated incompressibility at 4/11 thereby confirming the existence of an FQHE state at this filling. We note that the spin-polarization of the experimentally observed state at 4/11 has not yet been measured~\cite{Samkharadze15b, Pan15}. Although it is possible that the state is not fully-polarized at 4/11~\cite{Mukherjee14b, Balram15}, in this article, we only speculate on the nature of a fully-polarized state that could occur at this filling.

According to the CF theory, the $\nu=4/11$ state of electrons in the LLL is described by the $\nu^{*}=4/3=1+1/3$ state of CFs [electronic filling factor $\nu$ is related to the CF filling factor $\nu^{*}$ by $\nu=\nu^{*}/(2\nu^{*}+1)$]. The natural assumption that CFs form a Laughlin state in their second CF-Landau-like level [called Lambda level ($\Lambda$L)] is not supported by numerical calculations~\cite{Mandal02, Mukherjee14}. Studies of the pseudopotentials $\{V_{m}\}$ ($V_{m}$ denotes the pseudopotential of relative angular momentum $m$, which is the energy cost of placing two particles in the relative angular momentum $m$ state) of CFs~\cite{Sitko96,Wojs00,Lee01,Lee02,Wojs04,Balram16c,Balram17b} show that the interaction between CFs in their second $\Lambda$ level (S$\Lambda$L) is strongly repulsive in the relative angular momentum \emph{three} channel. For comparison, the electron-electron interaction in the lowest Landau level, which stabilizes the conventional Laughlin state at $\nu=1/3$~\cite{Laughlin83}, has the strongest repulsion in the relative angular momentum \emph{one} channel.  Motivated by these observations, W\'ojs, Yi, and Quinn (WYQ) suggested using the $V_{3}$-only Hamiltonian to describe the physics of interacting CFs residing in their S$\Lambda$L~\cite{Wojs04}. 

Despite extensive numerical work, it has not yet been possible to make a definitive statement on whether the WYQ Hamiltonian produces a gapped ground state at $\nu=1/3$. In particular,  the nature of the ground state produced by the WYQ Hamiltonian has a strong geometry dependence.  For all finite systems amenable to exact diagonalization in the spherical geometry, the ground state of the WYQ Hamiltonian is uniform (and topologically distinct from the $1/3$ Laughlin state) and well separated from other excited states~\cite{Jolicoeur17a}. Moreover, for all systems considered, the quantum numbers of the quasiparticle and quasihole, obtained by removing and inserting a single flux quantum into the ground state, respectively are identical to those of the corresponding excitations of the $1/3$ Laughlin state.  Nonetheless, the gaps of the WYQ Hamiltonian seem to extrapolate to a nonpositive value in the thermodynamic limit, indicating that its ground state may be compressible~\cite{Jolicoeur17a}. In fact, on the torus and cylindrical geometries, the ground state of the WYQ Hamiltonian does appear to be in a bubble phase~\cite{Regnault16, Jolicoeur17a}. Thus, the physical origin and the precise nature of the $1/3$ FQHE state of CFs in their S$\Lambda$L has been controversial~\cite{Mandal02, Mukherjee14, Das20, Balram15}. We mention here that all these calculations simulate only the CFs in their S$\Lambda$L and ignore their lowest $\Lambda$L.  Thus strictly speaking, these results do not preclude incompressibility at $4/11$ for the WYQ model. Moreover, it is also possible that including the other pseudopotentials could stabilize an incompressible state.

In this work, we take a different approach and ask if a wave function can be directly written down for the $4/11$ state of electrons? We answer this question in the affirmative by constructing a parton state~\cite{Jain89b} at this filling. We show that the wave function of the parton state provides a viable representation of the actual Coulomb ground state in the LLL obtained in numerics. Furthermore, we work out in detail the low-energy effective theory of its edge and make predictions that could be tested out in experiments. At the moment the connection of our parton ansatz to the state of interacting CFs in their S$\Lambda$L is unclear.  An interpretation of the $4\bar{2}1$ state as the $\nu^{*}=1+1/3$ state of CFs could then, by particle-hole conjugation of CFs in their S$\Lambda$L~\cite{Balram17b}, also explain the 5/13 FQHE which corresponds to $\nu^{*}=1+2/3$. We note here that parton wave functions can represent states of interacting CFs. As a case in point, parton states that describe the pairing of CFs have recently been constructed~\cite{Wu17, Balram18, Bandyopadhyay18, Faugno19}. 

The article is organized as follows. In the next section we provide a background on parton states and introduce our parton ansatz for 4/11. Then in Sec.~\ref{sec: ED_comparison}, we compare our parton wave function with the exact lowest Landau-level Coulomb ground state. In Sec.~\ref{sec:eff_edge} we derive the low-energy effective theory of the parton edge. We close the paper in Sec.~\ref{sec: discussions} with a discussion of the experimental consequences of the parton ansatz.

\section{Parton states and the $4\bar{2}1^{3}$ ansatz}
\label{sec: parton_ansatz_4_11}
The parton theory~\cite{Jain89b} generalizes the Jain CF states~\cite{Jain89,Jain07} to a broader class of wave functions. In the parton theory one envisages breaking electrons into $k$ sub-particles called partons. The partons have filling factors $\nu_\delta$, where $\delta=1,2,\cdots, k$ labels the parton species. When each parton species resides in an IQHE state, i.e. $\nu_{\delta}$ is an integer for all $\delta$, then an incompressible state can be achieved.  The resulting wave functions, labeled ``$n_1n_2n_3...$," are given by
\begin{equation}
\Psi^{n_1n_2n_3...}_\nu = \mathcal{P}_{\rm LLL} \prod_{\delta=1}^{k}\Phi_{n_\delta}(\{z_j\}),
\label{eq:parton_wf}
\end{equation}
where $z_{j}=x_{j}-iy_{j}$ is the two-dimensional coordinate of the $j$th electron parametrized as a complex number, $\Phi_n$ is the Slater determinant wave function for the state with $n$ filled Landau levels of electrons, and $\mathcal{P}_{\rm LLL}$ denotes projection into the LLL, as is appropriate for the large-magnetic-field limit. 

Since the partons are unphysical, they should be glued back together into the physical electrons. This operation, at the level of wave functions, is equivalent to setting the parton coordinates $z_j^\delta$ equal to the parent electron coordinates $z_j$, i.e., $z_j^\delta = z_j$ for all $\delta$.  Thus, each of the IQHE states in the wave function of Eq.~(\ref{eq:parton_wf}) is made up of \emph{all} the electrons (denoted by $\{z_j\}$). The partons can experience effective magnetic fields opposite to that of the field experienced by electrons. These correspond to negative filling factors for the partons, which we denote as $\bar{n}$, with $\Phi_{\bar{n}}=\Phi_n^*=\Phi_{-n}$.  Since each parton species occurs at the same density as that of the electrons, the charge of the $\delta$ parton species is given by $e_\delta = -\nu e / \nu_\delta$. The constraint $\sum_{\delta=1}^{k} e_\delta=-e$ implies that the electronic filling factor is related to the partonic filling factor by $\nu = [\sum_{\delta=1}^{k} \nu_\delta^{-1}]^{-1}$.  To describe a fermionic (bosonic) state, the number of parton species $k$ has to be odd (even). 

Many notable classes of FQHE states can be reinterpreted as parton states. The Laughlin state~\cite{Laughlin83} at $\nu=1/p$, described by the wave function $\Psi_{\nu=1/p}^{\rm Laughlin}=\Phi^{p}_{1}$, can be interpreted as the $\underbrace{11\cdots}_{p}$ parton state. The Jain states, described by the wave function $\Psi_{\nu=n/(pn\pm 1)}^{\rm Jain}=\mathcal{P}_{\rm LLL}\Phi_{\pm n}\Phi^{p}_{1}$, appear as the ${\pm n}\underbrace{11\cdots}_{p}$ states. Several parton states have recently been shown to be feasible candidates to describe FQHE states arising in the second LL of GaAs, wide quantum wells,  and in LLs of graphene~\cite{Wu17, Balram18, Bandyopadhyay18, Balram18a, Faugno19, Balram19, Kim19, Balram20, Balram20b, Faugno20a, Balram20a, Faugno20b}. These states often have exotic properties such as hosting excitations possessing non-Abelian braid statistics~\cite{Wen91}.  

In this article we propose the parton state denoted as ``$4\bar{2}1^{3}$" and described by the wave function
\begin{equation}
\Psi^{4\bar{2}1^{3}}_{4/11} = \mathcal{P}_{\rm LLL} \Phi_{4}[\Phi_{2}]^{*}\Phi^{3}_{1} \sim \frac{\Psi^{\rm Jain}_{4/9}\Psi^{\rm Jain}_{2/3}}{\Phi_{1}}
\label{eq: parton_4_11_4bar2111}
\end{equation}
for the 4/11 FQHE.  The $\sim$ sign in the above equation indicates that the states on either side of the sign differ in the precise details of how the projection to the LLL is carried out. We expect that such details do not change the topological phase of the wave function~\cite{Balram16b}. This state occurs at a shift~\cite{Wen92} of $\mathcal{S}=5$ on the spherical geometry. This is the same shift as that of the 4/11 state, which results from the formation of the unconventional 1/3 WYQ state in the S$\Lambda$L~\cite{Mukherjee14}.  Furthermore, it is the only known shift where the LLL Coulomb ground-state beyond a certain number of electrons is consistently uniform and incompressible for all finite systems that are accessible to numerics~\cite{Mukherjee14}. 

Using the Jain-Kamilla method~\cite{Jain97b, Moller05, Davenport12, Balram15a}, we can evaluate the Jain wave functions in \emph{real space} (in first quantized form) for fairly large system sizes using the Monte Carlo method.  This allows us to construct the form of the $4\bar{2}1^{3}$ state given on the rightmost side of Eq.~(\ref{eq: parton_4_11_4bar2111}) for large systems and thus, this is the form we shall use below. However, it is hard to obtain the Fock-space space (second-quantized) representation of our parton state. 

\section{Numerical comparisons with exact diagonalization results}
\label{sec: ED_comparison}
We shall use Haldane's spherical geometry~\cite{Haldane83} for all our numerical calculations. In the spherical geometry, $N$ electrons reside on the surface of a sphere and experience a radial magnetic field generated by a magnetic monopole placed at the center of the sphere. The strength of the magnetic monopole is denoted as $2Q$, where $2Q$ is a positive integer, and the total flux emanating from the sphere is given as $2Q(hc/e)$.  The state with $n$ filled Landau levels on the sphere can only be constructed when $N$ is divisible by $n$ and $N\geq n^2$. Thus, the $4\bar{2}1^{3}$ state can only be constructed for $N=4m$ with $N\geq 16$. 

The only system accessible to exact diagonalization is the smallest one with $N=16$ at $2Q=39$, which has a Hilbert-space dimension of about 683 million. Unfortunately, this system aliases with the $\bar{4}\bar{2}1^{3}$ state~\cite{Balram20b}. Furthermore,  the wave functions of the $4\bar{2}1^{3}$ and $\bar{4}\bar{2}1^{3}$ states are also identical for $N=16$ electrons. Nevertheless, since this is the only system for which we have exact results available, we shall go ahead and test the proposed $4\bar{2}1^{3}$ ansatz against it. The next system of $N=20$ electrons at $2Q=50$ has a Hilbert-space dimension of over 591 billion and is well beyond the reach of exact diagonalization. 

The exact LLL Coulomb spectrum of the system of $N=16$ electrons at flux $2Q=39$ was calculated in Ref.~\cite{Mukherjee14}. The LLL Coulomb ground state for this system is uniform on the sphere and has an excitation gap of about 0.002 $e^{2}/(\epsilon\ell)$~\cite{Mukherjee14}, where $\ell=\sqrt{\hbar c/(eB)}$ is the magnetic length and $\epsilon$ is the dielectric constant of the background host. The exact LLL Coulomb ground-state energy is $-0.43832~e^{2}/(\epsilon\ell)$ while the energy of the $4\bar{2}1^{3}$ state is $-0.4287(7)~ e^{2}/(\epsilon\ell)$ (the number in the parenthesis indicates the statistical uncertainty of the Monte Carlo estimate), which is within $2.2\%$ of the exact energy. Compared to the agreement between the exact LLL Coulomb ground states and the Laughlin and Jain states, this discrepancy is large. However, this discrepancy in energies is similar to that of candidate wave functions and exact Coulomb ground states in the second LL~\cite{Balram20b}. We also note that, although the $4\bar{2}1^{3}$ state does not provide a very accurate microscopic description of the LLL Coulomb ground state, it could potentially be in the same universality class as the actual ground state.

\begin{figure}[htpb]
\begin{center}
\includegraphics[width=0.47\textwidth,height=0.23\textwidth]{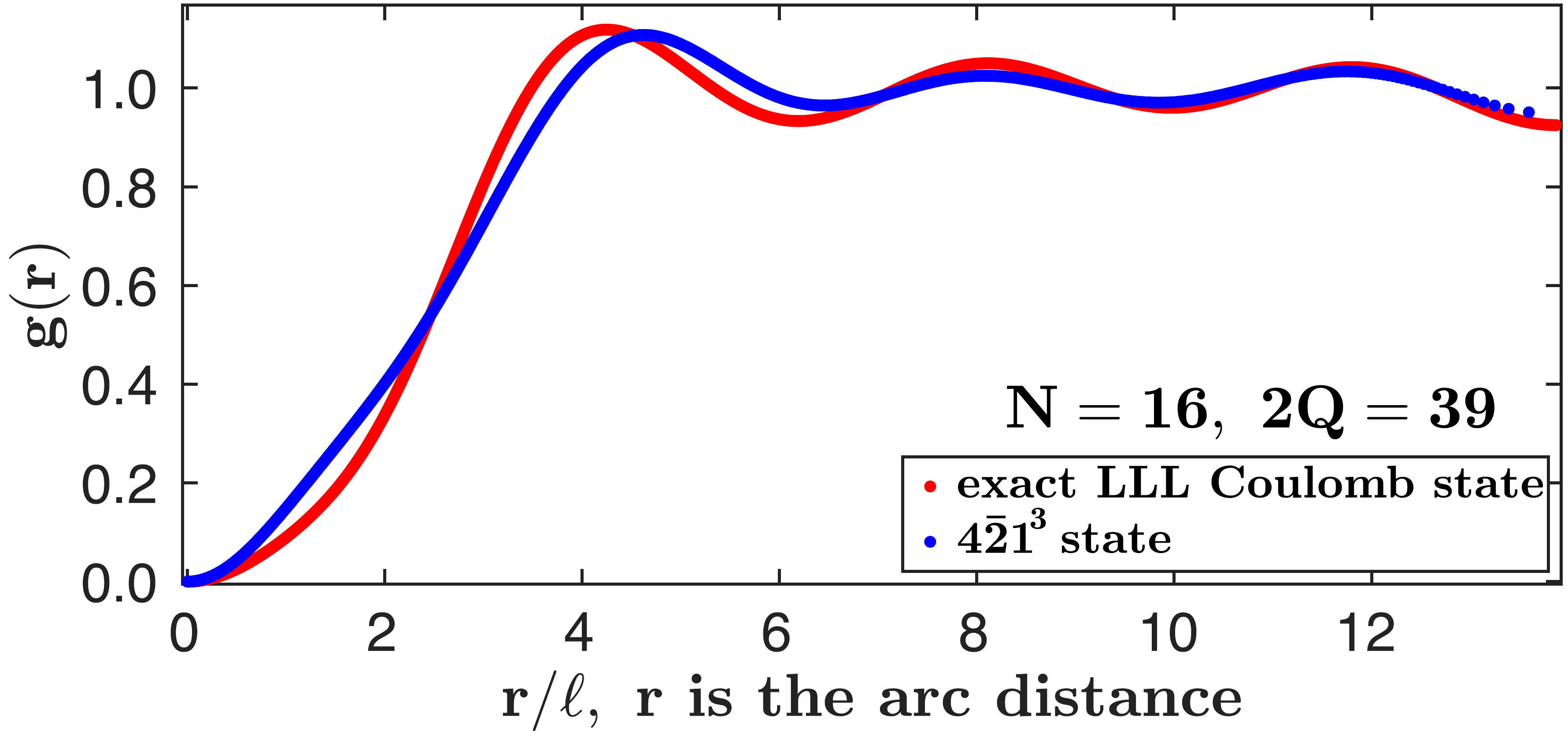} 
\caption{(color online) The pair-correlation function $g(r)$ as a function of the arc distance $r$ on the sphere for the exact lowest-Landau-level Coulomb ground state (red filled dots), and the $4\bar{2}1^{3}$ state of Eq.~(\ref{eq: parton_4_11_4bar2111}) (blue dots) for $N=16$ electrons at flux $2Q=39$.}
\label{fig: pair_correlations_4_11}
\end{center}
\end{figure}

The large Hilbert-space dimension of the system of $N=16$ electrons at flux $2Q=39$ precludes a calculation of the overlap of the parton wave function with the exact Coulomb ground state. Nevertheless, in Fig.~\ref{fig: pair_correlations_4_11} we have shown a comparison of the pair-correlation function $g(r)$ of these two states. The $g(r)$ of the parton and exact states are in reasonable agreement with each other. Both states show oscillations that decay at large distances which is a typical characteristic of incompressible states~\cite{Kamilla97, Balram15b, Balram17}. In comparison with the agreement of the Laughlin and Jain states with the exact LLL Coulomb ground state, the $g(r)$ of the exact and parton states differ slightly at the short-to-intermediate distances.

\section{Effective theory of the $4\bar{2}1^{3}$ edge}
\label{sec:eff_edge}
In this section we derive the low-energy effective theory of the $4\bar{2}1^{3}$ edge. To do so, we closely follow the procedure outlined in the Supplemental Material of Refs.~\cite{Balram18a, Balram20b}. For a major part of this section, we shall consider the $n\bar{2}1^{3}$ state with an arbitrary $n$ and then towards the end specialize to the case of $n=4$. The unprojected wave function of the $n\bar{2}1^{3}$ state can be re-written as
\begin{equation}
\Psi^{n\bar{2}1^{3}}_{2n/(5n+2)} =  \Phi_{n}[\Phi_{2}]^{*}\Phi^{3}_{1} = \Phi_{n}[\Phi_{2}]^{*}\Psi^{{\rm Laughlin}}_{1/3}.
\label{eq:parton_unprojected_nbar2111}
\end{equation}
This state can be expressed in terms of three partons $\wp = \wp_{1}\wp_{2}\wp_{3}$, where the $\wp_{i}$'s are in the following mean-field states: 
\begin{itemize}
 \item (a) $\wp_{1}$ is in a $\nu=n$ IQHE state,
 \item (b) $\wp_{2}$ is in a $\nu=-2$ IQHE state,
 \item (c) $\wp_{3}$ is in a $\nu=1/3$ Laughlin state.
\end{itemize}
The charges of these partons are $q_{1}=-2e/(5n+2)$, $q_{2}=ne/(5n+2)$ and $q_{3}=-6ne/(5n+2)$. The parton wave function of Eq.~(\ref{eq:parton_unprojected_nbar2111}) describes an Abelian state with a residual $U(1)\times U(1)$ gauge symmetry associated with the transformations:
\begin{equation}
\wp_{1}\rightarrow e^{i\theta_{1}} \wp_{1},~\wp_{2}\rightarrow e^{-i\theta_{1}+i\theta_{2}} \wp_{2},~
\wp_{3}\rightarrow e^{-i\theta_{2}} \wp_{3}.
\end{equation}
Therefore we have two internal emergent $U(1)$ gauge fields, denoted by $h_{\mu}$ and $g_{\mu}$, associated with $\theta_{1}$ and $\theta_{2}$ in the above transformations. The low-energy effective field theory for this parton mean-field state is described by the Chern-Simons Lagrangian density (henceforth in this section we set $e=h=c=1$ for convenience)~\cite{Balram18a}:
\begin{eqnarray}
\mathcal{L}&=&-\frac{1}{4\pi} \sum_{i=1}^{n}\alpha^{(i)}\partial \alpha^{(i)}+\frac{1}{2\pi}\sum_{i=1}^{n}(h-q_{1}A)\partial \alpha^{(i)}  \nonumber \\
&+& \frac{1}{4\pi} \sum_{j=1}^{2}\beta^{(j)}\partial \beta^{(j)}+\frac{1}{2\pi}\sum_{j=1}^{2}(g-h-q_{2}A)\partial \beta^{(j)}  \nonumber \\
&-& \frac{3}{4\pi} \gamma \partial \gamma+\frac{1}{2\pi}(-g-q_{3}A)\partial \gamma,
\label{eq:Lagrangian_density}
\end{eqnarray}
where $A$ is the external physical electromagnetic vector potential, and $\alpha^{(i)}$, $\beta^{(j)}$ and $\gamma$ are the $U(1)$ gauge fields describing the current fluctuations of the $\nu=n$, $\nu=-2$ IQHE and $\nu=1/3$ Laughlin states respectively. Furthermore, we have used the standard shorthand notation for the Chern-Simons term, i.e. $\alpha\partial \beta\equiv\epsilon^{\mu\nu\lambda}\alpha_{\mu}\partial_{\nu}\beta_{\lambda}$, where $\epsilon^{\mu\nu\lambda}$ is the fully antisymmetric Levi-Civita tensor and the Einstein summation convention is used (repeated indices are summed over).  The Lagrangian density of Eq.~(\ref{eq:Lagrangian_density}) describes a $U(1)^{n+3}$ Chern-Simons theory, where we have $n$ fields from the factor of $\Phi_{n}$, two from $\Phi_{-2}$ and one from $\Phi^{3}_{1}$.  \\

This Chern-Simons theory can be further simplified by integrating out the internal gauge fields $h$ and $g$. By doing so we obtain the following two constraints~\cite{Balram18a}
\begin{equation}
\sum_{i=1}^{n} \alpha^{(i)} = \sum_{j=1}^{2} \beta^{(j)} + c, 
\label{eq:constraint1}
\end{equation}
and
\begin{equation}
\gamma = \sum_{j=1}^{2} \beta^{(j)} + d,
\label{eq:constraint2}
\end{equation}
where $c$ and $d$ are $U(1)$ gauge fields that satisfy $\epsilon^{\mu\nu\lambda} \partial_\nu c_\lambda=0$ and $\epsilon^{\mu\nu\lambda} \partial_\nu d_\lambda=0$. 
Substituting Eqs.~(\ref{eq:constraint1}) and (\ref{eq:constraint2}) into Eq.~(\ref{eq:Lagrangian_density}) leads to a Lagrangian density in which all terms involving the gauge fields $c$ and $d$ vanish. Thus we end up with a simplified $U(1)^{n+1}$ Chern-Simons theory, which can be described by an integer valued symmetric $(n+1)\times (n+1)$ $K$ matrix~\cite{Balram18a}.

The constraints of Eqs.~(\ref{eq:constraint1}) and (\ref{eq:constraint2}) allow us to eliminate $\alpha^{(n)}$ and $\beta^{(2)}$. To do so we substitute
\begin{eqnarray}
\alpha^{(n)}&=&\gamma - \alpha^{(1)} - \alpha^{(2)} - \cdots - \alpha^{(n-1)}  +c -d, ~{\rm and}\nonumber \\
\beta^{(2)} &=&\gamma - \beta^{(1)} - d
\end{eqnarray}
back into the Lagrangian density given in Eq.~(\ref{eq:Lagrangian_density}), and use the fact that any term involving the gauge fields $c$ and $d$ vanish, to obtain the following simplified Lagrangian density:
\begin{widetext}
{\small
\begin{eqnarray}
\mathcal{L}&=&-\frac{1}{4\pi}\left( \alpha^{(1)}\partial \alpha^{(1)}+\alpha^{(2)}\partial \alpha^{(2)}+\cdots+\alpha^{(n-1)}\partial \alpha^{(n-1)}+
(\gamma - \alpha^{(1)} -\alpha^{(2)}-\cdots-\alpha^{(n-1)}) \partial (\gamma - \alpha^{(1)} -\alpha^{(2)}-\cdots-\alpha^{(n-1)}) \right)  \nonumber \\
&+& \frac{1}{4\pi}\left( \beta^{(1)}\partial \beta^{(1)} + (\gamma-\beta^{(1)})\partial (\gamma-\beta^{(1)})\right) \\
&-& \frac{3}{4\pi} \gamma \partial \gamma + \frac{1}{2\pi}A\partial \gamma.
\label{eq:Lagrangian_density_simplified}
\nonumber 
\end{eqnarray}
}
\end{widetext}
To put the Lagrangian density into its usual form~\cite{Wen91b,Wen92b,Wen95,Moore98} [see Eq.~(\ref{eq:Lagrangian_density_Kmatrix_charge_vector})], we define a new set of gauge fields:
\begin{equation}
 (a^{1},\cdots,a^{n+1}) = (\alpha^{(1)},\alpha^{(2)},\cdots,\alpha^{(n-1)},\beta^{(1)},\gamma),
\end{equation}
to obtain:
\begin{equation}
 \mathcal{L} = -\frac{1}{4\pi} K_{\rm IJ}a^{\rm I}\partial a^{\rm J} + \frac{1}{2\pi} t^{\rm I}A\partial a^{\rm I}.
\label{eq:Lagrangian_density_Kmatrix_charge_vector}
\end{equation}
Here the $K$ matrix is given by
\begin{equation}
K =   
\begin{pmatrix} 
       2 & 1 & 1\cdots1 & 0& -1\\
       1 & 2 & 1\cdots1 & 0& -1\\
       \vdots  & \ddots & \cdots & 0 &-1\\
       1 & \vdots &2 & 0 & -1 \\
        0 & \cdots & 0 & -2  & 1 \\
        -1 & \cdots & -1 & 1 &  3 \\
   \end{pmatrix},
\label{eq:Kmatrix_parton_nbar2111}   
\end{equation}
and the charge vector is $t=(0,0,\cdots,0,1)^{\rm T}$. The $K$ matrix $\{K_{ij}\}$ can be specified as
\begin{equation}
K_{i,j}=\begin{cases}
2, & 1\leq i=j \leq n \\
-2, & i=j=n \\
3, & i=j=n+1 \\
-1, & (i=n+1 \land j\leq n-1) \\
-1, & (i\leq n-1 \land j=n+1) \\
1, & 1\leq i<j\leq n-1 \\
1, & 1\leq j<i\leq n-1 \\
1, & (i=n \land j= n+1) \\
1, & (i=n+1 \land j= n)  \\
0, & {\rm otherwise} 
\end{cases}.
\end{equation}
The filling factor is related to the $K$ matrix as~\cite{Wen95}:
\begin{equation}
\nu =  t^{\rm T}\cdot K^{-1} \cdot t=K^{-1}_{n+1,n+1} = 2n/(5n+2).
\end{equation}
This value is consistent with the value ascertained from the microscopic wave function given in Eq~(\ref{eq:parton_unprojected_nbar2111}). The ground-state degeneracy of the $n\bar{2}1^{3}$ state on a manifold with genus $g$ is~\cite{Wen95}:
\begin{equation}
 \text{ground-state degeneracy} = |{\rm det}(K)|^{g} =(5n+2)^{g}.
\end{equation}
An interesting aspect to note is that when $n=2l$ is even, then the filling factor $\nu=2l/(5l+1)$. For these fillings, the ground-state degeneracy is $2^g (5l+1)^g$. Thus for even $n$ we get a single-component Abelian state at $\nu=r/q$ (with $r,q$ coprime), which has a ground-state degeneracy on the torus ($=2q$) that is greater than the denominator $q$ (see Refs.~\cite{Balram20,Balram20b,Faugno20b} for other examples of such states). 

The $K$ matrix of Eq.~(\ref{eq:Kmatrix_parton_nbar2111}) has one negative and $n$ positive eigenvalues which implies that the $n\bar{2}1^{3}$ state hosts $n$ forward-moving and one backward-moving edge modes, resulting in a chiral central charge of $n-1$. The charges and the braiding statistics of the quasiparticles of the $n\bar{2}1^{3}$ state can be determined from the above $K$ matrix following the work of Ref.~\cite{Wen95}. 

\subsubsection{Coupling to curvature and shift~\cite{Wen92}}
To compute the shift $\mathcal{S}$~\cite{Wen92} of the $n\bar{2}1^{3}$ state on the sphere, we need to couple the state to the curvature. For a state filling the $n^{\rm th}$ Landau level and described by the $U(1)$ gauge field $\zeta$, coupling to curvature results in an additional term in the Lagrangian density given by~\cite{Wen92}
\begin{equation}
\Delta\mathcal{L} = \frac{1}{2\pi} s~\omega \partial \zeta,
\end{equation}
where $\omega$ is the spin connection, and the spin $s=(n-1/2)$. Including the coupling to curvature for the $n\bar{2}1^{3}$ state, we get the following additional terms in the Lagrangian density:
\begin{equation}
\Delta\mathcal{L} = \frac{1}{2\pi} \sum_{i=1}^{n} (i-1/2)\omega \partial \alpha^{(i)} 
-\frac{1}{2\pi} \sum_{j=1}^{2} (j-1/2)\omega \partial \beta^{(j)} 
+ \frac{1}{2\pi} \frac{3}{2} \omega \partial \gamma. 
\end{equation}
Using the constraints of Eqs.~(\ref{eq:constraint1}) and (\ref{eq:constraint2}) we again get rid of $\alpha^{(n)}$ and $\beta^{(2)}$ to end up with the following additional term in the Lagrangian density that describes coupling of the $n\bar{2}1^{3}$ state to the curvature:
\begin{widetext}
\begin{eqnarray}
\Delta\mathcal{L} &=& \frac{1}{2\pi} \left[ (1/2)\omega \partial \alpha^{(1)} + (3/2)\omega \partial \alpha^{(2)}+ \cdots + ((2n-3)/2)\omega \partial \alpha^{(n-2)}+
((2n-1)/2) \omega \partial (\gamma - \alpha^{(1)} -\alpha^{(2)}-\cdots-\alpha^{(n-1)})\right] \nonumber \\
&-&\frac{1}{2\pi} \left[ (1/2)\omega \partial \beta^{(1)} + (3/2)\omega \partial (\gamma - \beta^{(1)})\right]+ \frac{1}{2\pi} \frac{3}{2} \omega \partial \gamma=\frac{1}{2\pi} \mathfrak{s}^{\rm I}\omega\partial a^{\rm I}.
\end{eqnarray}
\end{widetext}
Here we have introduced the spin vector $\mathfrak{s}=\left[-(n-1),-(n-2),\cdots,-1,1,((2n-1)/2) \right]^{\rm T}$. The shift $\mathcal{S}$ is then given by~\cite{Wen95}
\begin{equation}
 \mathcal{S}=\frac{2}{\nu} \left( t^{\rm T}\cdot K^{-1} \cdot \mathfrak{s} \right)=1+n,
\end{equation}
which is consistent with the value derived from the microscopic wave function given in Eq.~(\ref{eq:parton_unprojected_nbar2111}).

\subsubsection{Specializing to the $n=4$ case for the $\nu=4/11$ state}
For the case of $n=4$ the charge vector $t$ and the $K$ matrix following Eq.~(\ref{eq:Kmatrix_parton_nbar2111}) are given by
\begin{equation}
K  =
\begin{pmatrix} 
      2 & 1 & 1 & 0 & -1 \\
      1 &  2 & 1 & 0 & -1 \\
      1 &  1 & 2 & 0 & -1 \\
       0 &  0 & 0 &-2 & 1 \\
       -1 & -1 &  -1 & 1 & 3\\
   \end{pmatrix}~{\rm and}~
   t= \begin{pmatrix} 
      0\\
      0\\
      0\\
      0\\
      1\\
   \end{pmatrix}.
\end{equation} 
The filling fraction is:
\begin{equation}
\nu =  t^{\rm T}\cdot K^{-1} \cdot t= 4/11,
\end{equation}
as anticipated. The ground-state degeneracy on a manifold with genus $g$ is:
\begin{equation}
 \text{ground-state degeneracy} = |{\rm det}(K)|^{g} =22^{g}.
\end{equation}
This $K$ matrix has one negative and four positive eigenvalues which indicates that the $4\bar{2}1^{3}$ state has four downstream and one upstream edge modes resulting in a chiral central charge of $3$.  We note here that the results of Ref.~\cite{Das20} suggest that the $4/11$ state hosts three downstream edge modes, which also results in a chiral central charge of 3  

\section{Discussion}
\label{sec: discussions}
In this section, we discuss various experimentally measurable properties of the $4\bar{2}1^{3}$ ansatz that can reveal its underlying topological order. The smallest quasiparticle, generated by creating a particle in the factor of $\Phi_{4}$, carries a charge of $-e/11$.  A single quasiparticle of charge $-2e/11$ and $-4e/11$ can be produced by creating a particle in the factor of $\Phi_{\bar{2}}$ (obtained by complex conjugating a hole in the factor of $\Phi_{2}$) or a particle in the factor of $\Phi_{1}$, respectively. All the excitations of the $4\bar{2}1^{3}$ state obey Abelian braiding statistics.

Owing to the factor of $\bar{2}$ the $4\bar{2}1^{3}$ state is expected to host backward moving modes which can be detected experimentally~\cite{Bid10, Dolev11}. \emph{Assuming} a complete equilibration of the edge modes, the thermal Hall conductance $\kappa_{xy}$ of the $4\bar{2}1^{3}$ state, at temperatures much smaller than the gap, is $\kappa_{xy}=3[\pi^2 k^{2}_{B}/(3h)T]$.  Recently, thermal Hall measurements have been carried out at many filling factors in the lowest Landau level~\cite{Banerjee17}. The Hall viscosity $\eta_{H}$ or the Lorentz shear modulus of the $4\bar{2}1^{3}$ state is also expected to be quantized~\cite{Read09}: $\eta_{H}=\hbar \rho_{{\rm 2D}} \mathcal{S}/4$, where $\rho_{{\rm 2D}}=(4/11)/(2\pi\ell^{2})$ is the electron density and $\mathcal{S}=5$ is the shift of the $4\bar{2}1^{3}$ ansatz.

The $4\bar{2}1^{3}$ state is the $n=4$ member of the $n\bar{2}1^{3}$ sequence, which produces states at $\nu=2n/(5n+2)$ with shift $\mathcal{S}=n+1$. The $n=1$ member produces the standard CF state at $\nu=2/7$. The $n=2$ member produces a novel $\mathbb{Z}_{2}$ state at filling factor $1/3$~\cite{Balram20}, which may be relevant for the second LL but is unlikely to be stabilized in the LLL~\cite{Balram20, Faugno20b}. The $3\bar{2}1^{3}$ state provides a candidate wave function at $\nu=6/17$, where some features of FQHE have been seen in the LLL~\cite{Pan03}, but as yet there is no definitive confirmation of incompressibility~\cite{Pan14, Samkharadze15b}. We discussed the $n=4$ member in detail in this work. The $n=5$ member produces a state at $\nu=10/27$, where no signs of FQHE have been reported to date. The $6\bar{2}1^{3}$ state occurs at $\nu=3/8$, where some features of FQHE have been observed, but incompressibility has not been conclusively established~\cite{Samkharadze15b}.

Finally, we mention the possibility of unpolarized states at 4/11. In the LLL, we expect the partially spin-polarized and spin-singlet $4/11$ states to arise from the analogous CF states at $\nu^{*}=4/3$~\cite{Park00b, Chang03b, Balram15}.  The $4\bar{2}1^{3}$ state also readily admits the possibility of unpolarized states building on the partially polarized states at $\nu=4$ and the spin-singlet states at $\nu=4$ and $\nu=2$. These states could potentially be relevant for some interactions.

\textit{Acknowledgements} - We acknowledge useful discussions with Maissam Barkeshli, Jainendra K. Jain, Thierry Jolicoeur, Sudhansu S. Mandal, and Arkadius\'z W\'ojs. Computational portions of this research work were conducted using the Nandadevi supercomputer, which is maintained and supported by the Institute of Mathematical Science's High-Performance Computing Center. We thank the Science and Engineering Research Board (SERB) of the Department of Science and Technology (DST) for funding support via the Startup Grant No. SRG/2020/000154.

\bibliography{../../Latex-Revtex-etc./biblio_fqhe}
\bibliographystyle{apsrev}
\end{document}